\newcommand{\ket}[1]{\ensuremath{\left|  #1 \right\rangle}}
\newcommand{\bra}[1]{\ensuremath{\left\langle  #1 \right|}}

\newcommand{\ind}[1]{_{\text{#1}}}

\documentclass[12pt,a4paper]{article}

\usepackage[normalem]{ulem}
\usepackage{amsmath,amssymb,graphicx,color,verbatim,soul}
\usepackage{times}

\definecolor{MyDarkBlue}{rgb}{0,0,0.75}
\usepackage[pdftex,pdfborder={0 0 0},citecolor=MyDarkBlue, urlcolor=MyDarkBlue,
  linkcolor=MyDarkBlue, colorlinks=true, bookmarksopen=true]{hyperref}

\usepackage{EPRKunkel_CiteStyle}

\newenvironment{sciabstract}{%
\begin{quote} \bf}
{\end{quote}}

\newcounter{lastnote}

\title{Spatially distributed multipartite entanglement enables\\ Einstein-Podolsky-Rosen steering of atomic clouds} 

\author
{Philipp Kunkel,$^{1\ast}$ Maximilian Pr\"ufer,$^{1}$ Helmut Strobel,$^{1}$ Daniel Linnemann,$^{1}$ \\Anika Fr\"olian,$^{1}$ Thomas Gasenzer,$^{1}$ Martin G\"arttner,$^{1}$ Markus K.\ Oberthaler$^{1}$\\
\\
\normalsize{$^{1}$Kirchhoff-Institut f\"ur Physik, Universit\"at Heidelberg,}\\
\normalsize{Im Neuenheimer Feld 227, 69120 Heidelberg, Germany}\\
\\
\normalsize{$^\ast$To whom correspondence should be addressed; E-mail:  steering@matterwave.de.}
}

\date{June 28, 2017}

\begin{document}
	
\renewcommand{\figurename}{\textbf{Fig.}}

\maketitle 

\begin{sciabstract}
	A key resource for distributed quantum-enhanced protocols is entanglement between spatially
separated modes. Yet, the robust generation and detection of nonlocal entanglement between spatially separated regions of an ultracold atomic system remains a challenge. Here, we use spin mixing in a tightly confined Bose-Einstein condensate to generate an entangled state of indistinguishable particles in a single spatial mode. We show experimentally
that this local entanglement can be spatially distributed by self-similar expansion
of the atomic cloud. Spatially resolved spin read-out is used to reveal a particularly strong form of quantum correlations known as Einstein-Podolsky-Rosen steering between distinct parts of the expanded cloud. Based on the strength of Einstein-Podolsky-Rosen steering we construct a witness, which testifies up to genuine five-partite entanglement.
\end{sciabstract}
The concept of quantum entanglement requires the definition of distinct physical subsystems.
For each subsystem, quantum mechanics poses a fundamental limit on the simultaneous knowledge of two non-commuting observables, $\hat{Q}_{\text A}$ and $\hat{P}_\text{A}$. 
This limit is given by the Heisenberg uncertainty relation for the variances of the observables $\Delta^2{Q}_\text{A}\Delta^2{P}_\text{A}\geq \left|\left\langle\left[\hat{Q}_\text{A},\hat{P}_\text{A}\right]\right\rangle\right|^{2}/4$. 
Einstein, Podolsky and Rosen (EPR) pointed out that quantum mechanics allows for nonlocal correlations between two separate systems A and B which are at odds with the assumptions of local realism \cite{Einstein1935}.
As a reaction, Schr\"odinger argued that nonlocal EPR correlations enable what he called steering \cite{Schrodinger1935}.
This means that it is possible to infer from the measurement result obtained in system B the corresponding outcome in system A more accurately than allowed by the local uncertainty constraint \cite{Reid1989} (in the following phrased ``A steered by B'').
Steering is possible only if A and B are strongly entangled which renders it a witness for entanglement.
Originally intended to question the completeness of quantum mechanics, entanglement and nonlocality are now regarded as a resource for quantum technologies, such as quantum metrology~\cite{Giovannetti2004}, quantum cryptography \cite{Gisin2002}, and quantum information processing \cite{Braunstein2005}.

Pioneering work on nonlocal entanglement has been done in pure photonic systems and in hot atomic vapors building on atom light interaction \cite{Grangier1987,Ou1992,Julsgaard2001,Bachor2004,Cerf2007}. Ultracold atomic gases offer additional possibilities due to the high level of coherence of the internal as  well as the motional degrees of freedom.
Various schemes for generating nonlocal entanglement in the latter system have been discussed using quantum gate operations in optical lattices \cite{Bloch2008,Dai2016} and long-range interactions in Rydberg systems \cite{Zeiher2015,Labuhn2016}. In the continuous-variable limit considered in this work the generation via nonlinear dynamics of spatial multimode systems has been proposed  \cite{He2011,Bar2011,Kurkjian2013}.
Here we present a robust method to spatially distribute locally generated entanglement \cite{Peise2015,Pezze2016} in the spin degree of freedom of a Bose-Einstein condensate (BEC) by subsequent expansion of the atomic cloud. This constitutes an explicit experimental implementation of the recently formulated mapping of indistinguishable-particle entanglement in one mode to individually addressable subsystems \cite{Hyllus2012,Killoran2014}.

\begin{figure}
	\linespread{1}
	\centering
	\includegraphics[width = 0.5\columnwidth]{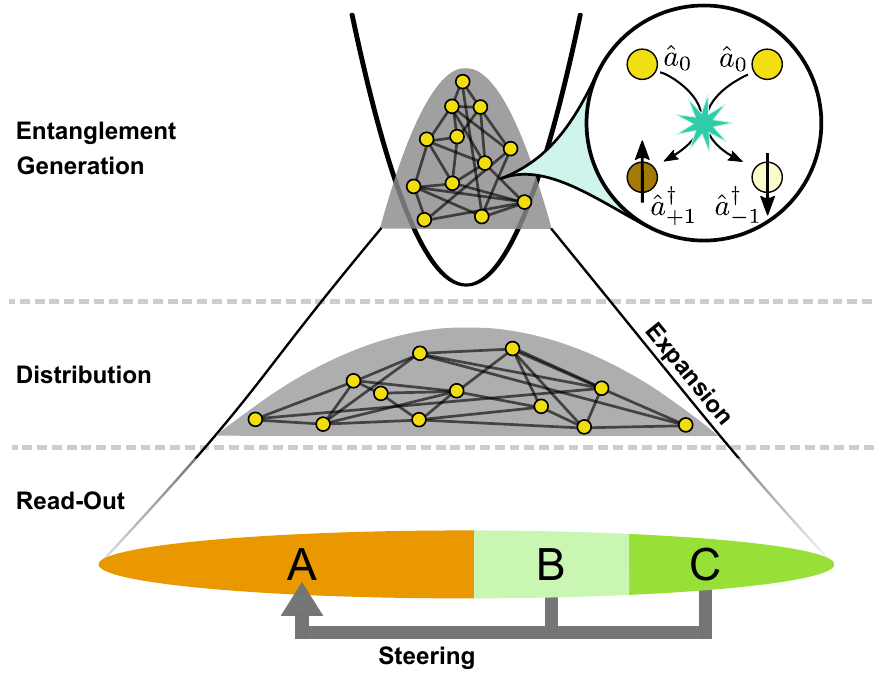} 
	\caption{\textbf{Distribution of entanglement.} In a tightly trapped Bose-Einstein condensate entanglement in the spin degree of freedom is generated by local spin-mixing interactions. Switching off the longitudinal confinement leads to a rapid expansion of the atomic cloud, which distributes the entanglement spatially.
		After local spin measurements with high spatial resolution we partition the detected atomic signal into distinct subsystems. 
		We demonstrate EPR steering between these parts, which evidences the presence of nonlocal quantum correlations and multipartite entanglement.}
	\label{Distribution}
\end{figure}
Experimentally, we prepare a BEC of $N\approx11,000$ $^{87}$Rb atoms in the $F = 1$ hyperfine manifold in the magnetic substate $m_{\text{F}} = 0$.
We initiate spin dynamics which coherently populates the states $m_{\text{F}} = \pm 1$ with correlated particle pairs \cite{Stamper2013}, leading to spin-nematic squeezing \cite{Hamley2012}.
This leads to entanglement shared among all atoms in the condensate.
Self-similar expansion for distributing the entanglement  is initiated by switching off the longitudinal confinement.
The expanding cloud evolves in the remaining waveguide potential.
After imaging with high optical resolution we analyze partitions of the resulting absorption signal to reveal entanglement and EPR steering between the corresponding atomic subsystems (see Fig.~\ref{Distribution}).

As non-commuting observables, $\hat Q$ and $\hat P$, we choose the spin operators $\hat{F}(0)$ and $\hat{F}(\pi/2)$, where $\hat{F}(\phi)= \left[(\hat{a}^\dagger_{+1}+\hat{a}^\dagger_{-1})\mathrm e^{i(\phi-\phi_0)}\hat{a}_{0} + \text{h.c.}\right]/\sqrt{2}$. Here, $\hat{a}^\dagger_{i}$ is the creation operator for a particle in the spin state $m_{\text{F}}=i$, h.c.~denotes the Hermitian  conjugate, and $\phi_0$ is an offset phase.
In the case of negligible populations in $m_\text{F}=\pm1$ compared to the total atom number $N$, which is fulfilled in the experiment, these operators obey the commutation relation \cite{SuppInfo}
\begin{equation}
	\left[\hat{F}(0),\hat{F}(\pi/2)\right] = 2i\hat{N}\,.
	\label{FComm}
\end{equation} 
By adjusting the hold time under the influence of the second-order Zeeman shift, the phase $\phi$ is precisely controlled.
To map $\hat{F}(\phi)$ on the detectable population difference $N^-(\phi) = N_{+1}-N_{-1}$ we apply a resonant radiofrequency pulse corresponding to a $\pi/2$ spin rotation after the expansion~\cite{SuppInfo}.

Because of the commutation relation~\eqref{FComm} the observed variances of the population differences after spin rotation fulfill the uncertainty relation
\begin{equation}
	\frac{\Delta^2 N^-(0)}{N}\frac{\Delta^2 N^-(\pi/2)}{N}\geq 1.
\end{equation}
This inequality also applies locally to any subsystem with corresponding particle number.

\begin{figure}
	\linespread{1}
	\centering
	\includegraphics[width = 0.85\textwidth]{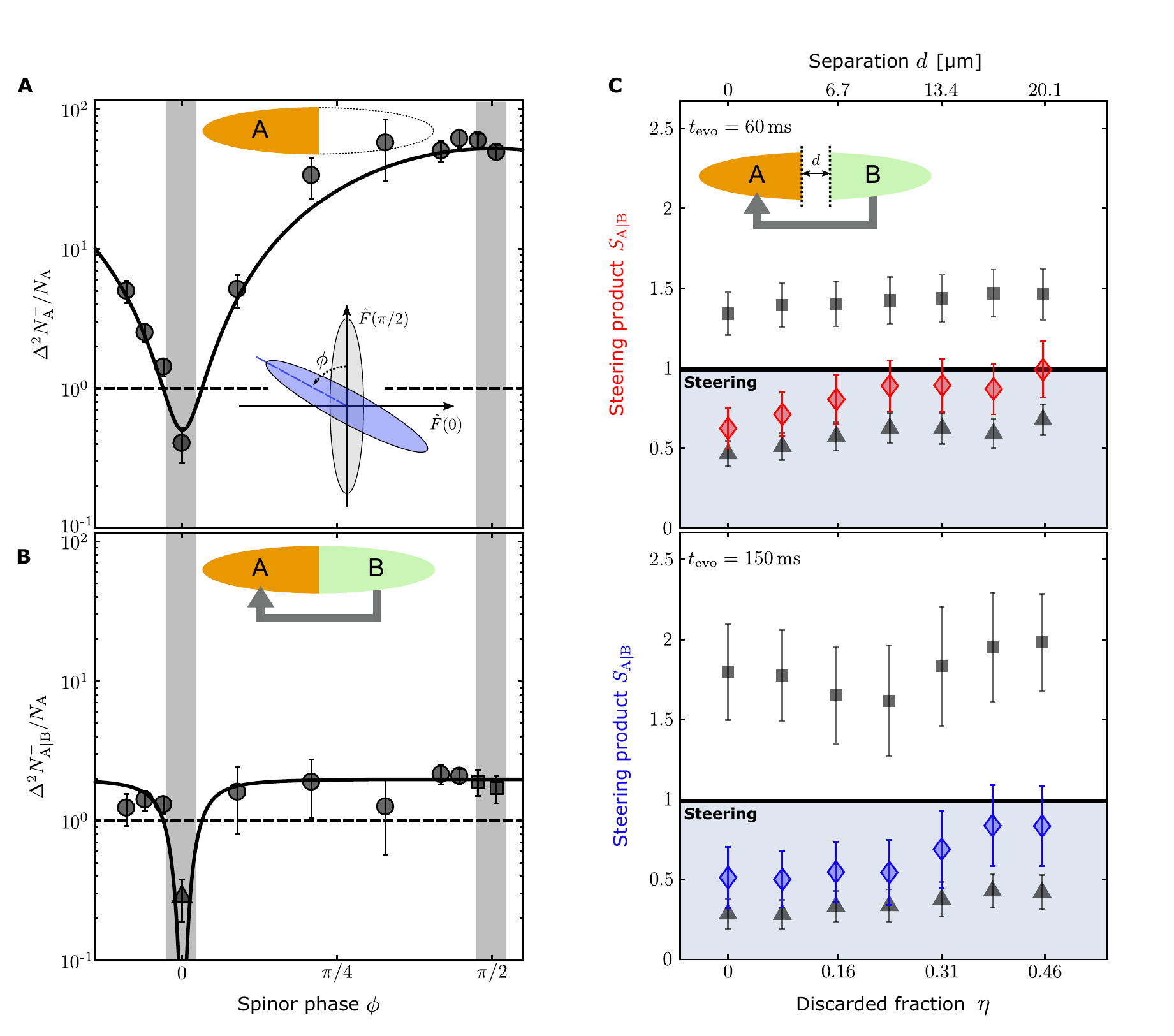}
	\caption{\textbf{Einstein-Podolsky-Rosen steering.} (\textbf{A}) A global change of the phase $\phi$ before the measurement allows a mapping of the spin observable $\hat F(\phi)$ to the read-out direction $\hat F(0)$ (inset). 
		Partitioning the atomic signal into two halves, we observe for subsystem A reduced and enhanced fluctuations of $\Delta^2 N^{-}_\text{A}/N_\text{A} = \Delta^2 (N_{\text{A},+1}-N_{\text{A},-1})/N_\text{A}$ compared to the shot-noise limit of a fully separable spin state (dashed line). 
		The solid line is a theoretical prediction based on our experimental parameters \cite{SuppInfo}. 
		At phase $\phi = 0$ one finds reduced fluctuations, while the fluctuations are enhanced at phase $\phi = \pi/2$. (\textbf{B}) The measurement result in B is used to infer the result in A (inset), leading to an inference variance $\Delta^2 N^{-}_{\text{A}|\text{B}}$. The solid line represents the theoretical prediction. The data in the gray shaded region are used to calculate the EPR steering product $S_{\text{A}|\text{B}}$.
		(\textbf{C}) We vary the spatial separation between the two subsystems by discarding a fraction $\eta$ of atomic signal in the middle of the cloud (inset). The red and blue diamonds are the products $S_{\text{A}|\text{B}}=\Delta^2 N^{-}_{\text{A}|\text{B}}(0)\Delta^2 N^{-}_{\text{A}|\text{B}}(\pi/2)/N_\text{A}^2$ of the inference variances after 60\,ms and 150\,ms of spin mixing time, respectively. The individual inference variances $\Delta^2 N^{-}_{\text{A}|\text{B}}/N_\text{A}$ at $\phi = 0$ and $\phi =\pi/2$ are shown as black triangles and squares, respectively. The steering product remains below the EPR steering bound even if a significant fraction of the atomic signal is discarded confirming the nonlocal character of the entanglement in our system. The given error bars correspond to an estimation of the 1 s.d. interval.}
	\label{Bipartite}
\end{figure}
In a first step we partition the absorption signal into two halves, A and B. 
In subsystem A we detect reduced (enhanced) fluctuations of $N_\text{A}^-(\phi)$ at phase $\phi = 0$ ($\phi=\pi/2$) as compared to the case of a fully separable initial state (see Fig.~\ref{Bipartite}A).
While the minimum fluctuations are below the separable-state limit (nematic squeezing), the variance product $\Delta^2 N^{-}_{\text{A}}(0)\Delta^2 N^{-}_{\text{A}}(\pi/2)/N^2_\text{A} = 0.41\cdot 49$ clearly exceeds the uncertainty limit.
To reveal EPR steering of A we demonstrate that a measurement in subsystem B can be used to infer the outcome in A with an accuracy beating the local uncertainty limit $\Delta^2 N^{-}_{\text{A}}(0)\Delta^2 N^{-}_{\text{A}}(\pi/2)/N^2_\text{A}\geq1$.

The outcome in A can be estimated by an arbitrary function of the measurement result in B \cite{Reid2009}. Here we construct an estimator based on five subdivisions of B.
The corresponding values of $N^{-}_{\text{B},k}$ are used to infer the result in A via the linear combination $N^{-}_{\text{A},\text{inf}}(\phi) = \sum_{k=1}^5 g_k(\phi) N^-_{\text{B},k}(\phi)$. The real numbers $g_k(\phi)$ are chosen to minimize the inference variance
\begin{equation}
		\Delta^2 N^{-}_{\text{A}|\text{B}}(\phi) = \Delta^2\left(N^{-}_{\text{A}}(\phi)-N^{-}_{\text{A},\text{inf}}(\phi)\right),		
\end{equation}
which is depicted in Fig.~\ref{Bipartite}B.
The inference variance quantifies the accuracy with which $N^-_\text{A}$ can be inferred by the estimator $N^-_\text{A,inf}$.
To compare the achieved accuracy with the local uncertainty relation we evaluate the steering product
\begin{equation}
	S_{\text{A}|\text{B}} = \frac{\Delta^2N^{-}_{\text{A}|\text{B}}(0)}{N_\text{A}}\frac{\Delta^2N^{-}_{\text{A}|\text{B}}(\pi/2)}{N_\text{A}}.
\end{equation}
$S_{\text{A}|\text{B}}< 1$ signals EPR steering of A by B.
In our experiment we obtain a value of $S_{\text{A}|\text{B}} = 0.62\pm0.12$ and $S_{\text{A}|\text{B}} = 0.51\pm0.19$ after 60\,ms and 150\,ms of spin mixing dynamics, respectively, verifying bipartite EPR steering in our system. 
The given errors correspond to the statistical estimation of one standard deviation applying a resampling method.
For all given variances, the independently characterized photon shot noise contribution to the absorption signal has been subtracted.
To underline the nonlocal aspect of steering we discard a fraction $\eta$ of the atoms in a region between A and B. Figure~\ref{Bipartite}C shows that EPR steering can be verified up to a discarded fraction of $\sim30\%$ of the atoms which corresponds to a minimal distance of $\sim 13\,\mu$m between the two systems.
This is consistent with monogamy of steering \cite{Reid2013}, which implies that by discarding more than a third of the whole system no steering between equal partitions of the remaining system is possible.

\begin{figure}[t!]
	\linespread{1}
	\centering
	\includegraphics[width = 0.5\columnwidth]{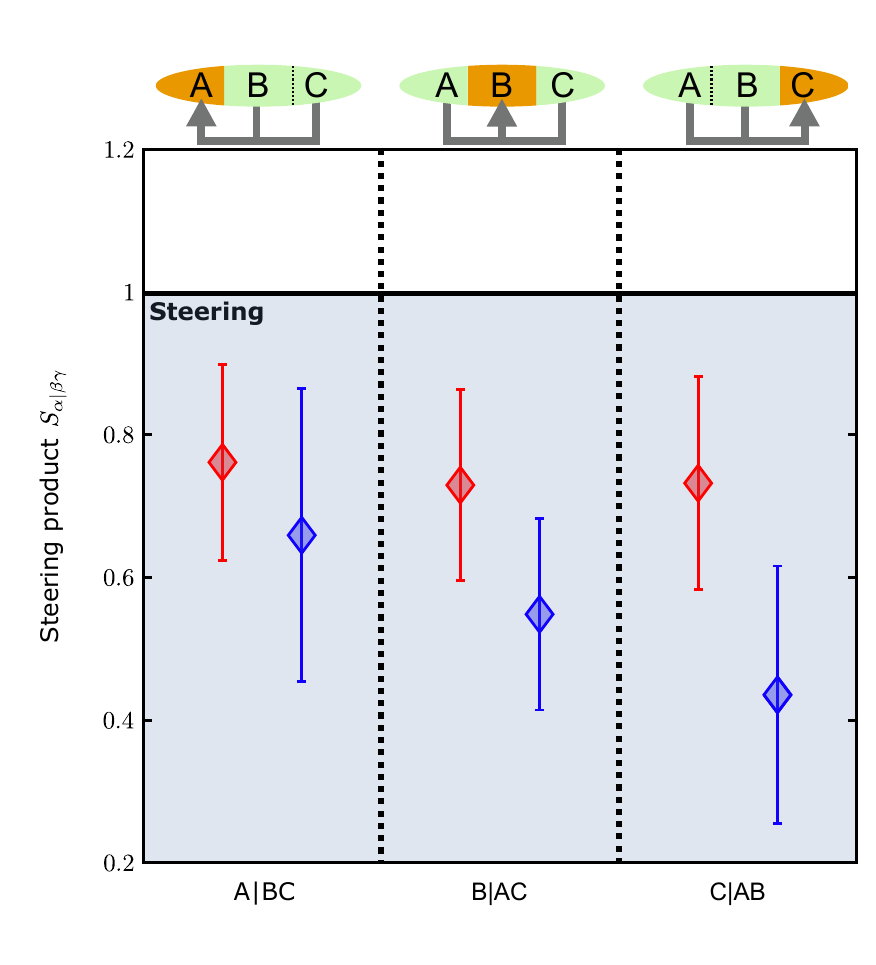}
	\caption{\textbf{Threeway EPR steering.} Partitioning the absorption signal into three parts of equal length ($\sim$ 20$\,\mu$m) we reveal that each of the three subsystems is steered by the other two. For each case, we calculate the steering product $S_{\alpha|\beta\gamma} = \Delta^2 N^{-}_{\alpha|\beta\gamma}(0)\Delta^2 N^{-}_{\alpha|\beta\gamma}(\pi/2)/N_\alpha^2$, where $N^{-}_{\alpha|\beta\gamma}(\phi)$ denotes the optimal inference on the observable $N^{-}_\alpha(\phi)$ in system $\alpha$ using the information obtained from the respective other two subsystems $(\beta,\gamma)$ (see text). The red (blue) points are the results for 60\,ms (150\,ms) of spin mixing time. The black line represents the steering bound. The given error bars correspond to an estimation of the 1 s.d. interval.}
	\label{Tripartite}
\end{figure}
For indistinguishable particles one expects that the entanglement is uniformly distributed over the whole system. 
We illustrate this by partitioning the absorption signal into three parts of equal length. 
Analogous to the previous discussion we evaluate the inference variance $\Delta^2N^{-}_{\text{A}|\text{BC}}/N_{\text{A}}$ for all permutations of ABC.  
Figure~\ref{Tripartite} summarizes that each part is steered by the remaining atomic cloud confirming threeway steering \cite{Armstrong2015}.
\begin{figure}[t!]
	\linespread{1}
	\centering
	\includegraphics[width = 0.5\columnwidth]{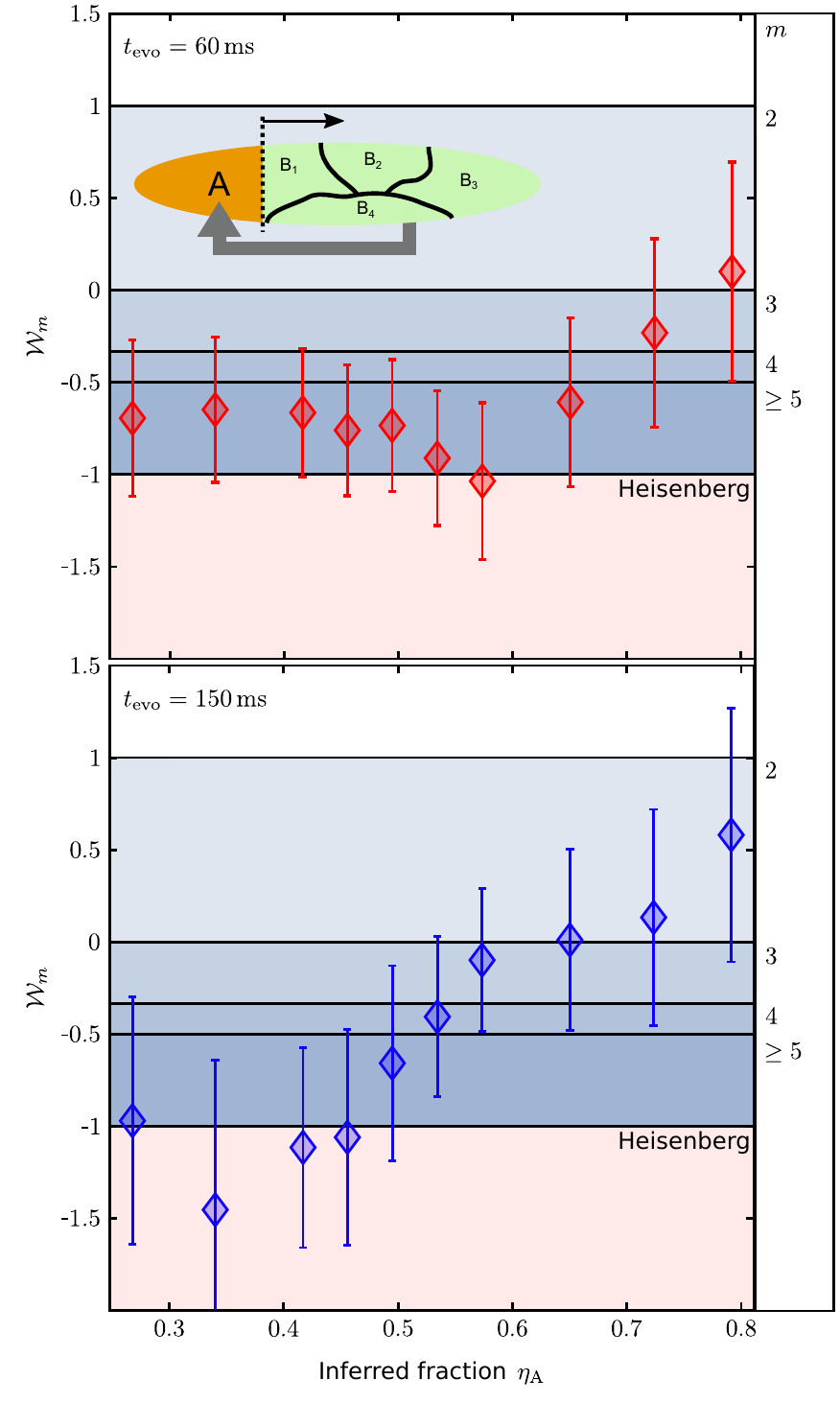}
	\caption{\textbf{Genuine multipartite entanglement.} In the bipartite steering scenario, the possible inference of system B on A is used to reveal genuine multipartite entanglement. For each partition A$|$B, quantified by $\eta_\text{A} = N_\text{A}/N$, system B can be divided into additional $m-1$ parties of equal atom number (see inset for an example). The regions where genuine $m$-partite entanglement is witnessed are indicated by the blue shadings, where the corresponding $m$ is given on the right. The upper (lower) panel shows the results for 60\,ms (150\,ms) of spin mixing time. 
		The lowest bound is given by the Heisenberg uncertainty limit for our observables in the full system. The given error bars correspond to an estimation of the 1 s.d. interval.}
	\label{Multipartite}
\end{figure}
It is important to note that for too small partitions spurious effects of the imaging technique become relevant.
Since the position of each atom is mapped onto a spatially distributed absorption signal, classical correlations are dominant below a certain length scale.
By analyzing a fully separable coherent spin state we confirm that for the partitions chosen here classical correlations are negligible~\cite{SuppInfo}.

The observation of EPR steering allows for statements about multipartite entanglement.
Specifically, the steering product can be used to construct a witness $\mathcal{W}_m$ for genuine $m$-partite entanglement. 
For this we partition the system into a subsystem A and the remainder B, which we divide into $m-1$ parts with equal atom numbers. 
Generalizing the derivation in \cite{Armstrong2015} we find that genuine $m$-partite entanglement \cite{Loock2003} is present if the inequality
\begin{equation}
	\mathcal{W}_m=\frac{\eta_\text{A}}{1-\eta_\text{A}}\,\frac{\left(1-\sqrt{S_{\text{A}|\text{B}}}\right)}{g(0)\,g(\pi/2)}<\frac{3-m}{m-1}
	\label{multientanglement}
\end{equation}
is fulfilled, given that $g(0)\,g(\pi/2)<0$~\cite{SuppInfo}.
Here $\eta_\text{A} = N_A/N$ denotes the fraction of atoms in system A, and the inferences are $N^{-}_{\text{A,inf}}(\phi) = g(\phi)N^{-}_{\text{B}}(\phi)$.
Fulfilling Eq.~\eqref{multientanglement} demonstrates that the quantum state of the system cannot be written as a mixture of states separable with respect to all possible bipartitions.
This implies that each part, or conjunction of parts, is entangled with the rest of the system.
Fulfilling Eq.~\eqref{multientanglement} in the limit $m\to \infty$ is excluded by the Heisenberg uncertainty limit of the full system.
Experimentally, we partition the absorption data of the atomic cloud into two parts and vary the fraction $\eta_\text{A}$ (inset Fig.~\ref{Multipartite}). In this way we verify up to genuine 5-partite entanglement (see Fig.~\ref{Multipartite}).

Our results combined with the well developed toolbox for the manipulation of ultracold gases give new perspectives for applications as well as fundamental questions.
Retrapping and storage of the produced states in tailored potentials enable quantum enhanced sensing of spatially varying external fields.
With the possibility of local control the deterministic generation of more general classes of nonlocal entangled states including cluster states, useful for continuous-variable quantum computation, is in reach \cite{Menicucci2006}.
Our general strategy for the detection of nonlocal entanglement can be applied to fundamental questions concerning the role of entanglement for long-time dynamics and thermalization of quantum many-particle systems \cite{Rigol2008}.

Complementary to our work, the group of P. Treutlein has detected spatial entanglement patterns, and the group of C. Klempt has observed entanglement of spatially separated modes.

\vspace{0.5cm}
\noindent\textbf{Acknowledgements}\\
We thank Margaret Reid and Philipp Hauke for discussions.\\
This work was supported by the Heidelberg Graduate School of Fundamental Physics, the Heidelberg Center for Quantum Dynamics, the European Commission, within the Horizon-2020 programme, through the FET-Proactive grant AQuS (Project No. 640800) and the ERC Advanced Grant EntangleGen (Project-ID 694561) as well as the DFG Collaborative Research Center SFB1225 (ISOQUANT).\\

\newpage

\bibliographystyle{EPRKunkel_BibStyle}
\bibliography{EPRKunkel_Bib}

\clearpage

\vspace*{1cm}
\begin{center}
  \textbf{\huge Supplementary Materials}
\end{center}

\renewcommand{\theequation}{S\arabic{equation}}
\setcounter{equation}{0}

\setcounter{figure}{0}
\renewcommand{\figurename}{\textbf{Fig.}}
\renewcommand{\thefigure}{S\arabic{figure}}

\section*{Materials and Methods}
\subsection*{Quantum state preparation}
The starting point of our experiments is a Bose-Einstein condensate in a crossed dipole trap with trapping frequencies $(\omega_\parallel,\omega_\perp) = 2\pi\times(51,286)\,\text{Hz}$ in the internal state $(F,m_\text{F}) =(1,0)$.
The subsequent spin mixing dynamics is described by the single-mode Hamiltonian $\hat{H}_\text{SM} = 2\lambda [\hat{a}_0^\dagger \hat{a}_0^\dagger \hat{a}_{+1} \hat{a}_{-1} + \hat{a}_{+1}^\dagger \hat{a}_{-1}^\dagger \hat{a}_0 \hat{a}_0] + [\lambda(2\hat{N}_0-1)+q](\hat{N}_{+1}+\hat{N}_{-1})$
where $\lambda$ is the coupling constant and the detuning $q = q_\text{z}+q_\text{ac}$ is the sum of the second-order Zeeman shift ($q_\text{z} \approx 2\pi\times 149\,\text{Hz}$ at a magnetic field of $1.44$\,G) and the ac-Zeeman shift used to control the spin mixing dynamics and the phase $\phi$. The coupling constant is $2\lambda N_0\approx -2\pi\times 2.5\,\text{Hz}$ for the atomic densities in our setup. To induce the ac-Zeeman shift we use a power stabilized microwave with resonant Rabi frequency of $\Omega \approx 2\pi\times 9.5\,\text{kHz}$ and $\delta \approx 2\pi\times 156\,\text{kHz}$ blue detuned with respect to the $(1,0)\leftrightarrow(2,0)$ transition \cite{Gerbier2006}. By tuning $q$ in this way, we observe a resonance in the $(1,\pm1)$ population after a fixed evolution time of 1\,s. We set $\delta$ to the center of this resonance feature. With higher atom numbers ($\sim 40000$), we also observe resonances for lower values of $q$ with a spacing of $\sim 2\pi\times 3.5\,$Hz. These correspond to excited states of the effective potential for $(1,\pm1)$, which is the combination of the dipole trap and the interaction with the (1,0) condensate \cite{Klempt2009}. These features are suppressed for the lower atom numbers chosen for the present experiment, which ensures that the pairs in $(1,\pm1)$ occupy the lowest mode with negligible population of the excited modes.

\subsection*{Spin read-out and detection}

We stop spin changing collisions by switching off the microwave which induces the ac-Zeeman shift.
The resulting energy splitting between the states $(1,0)$ and $(1,\pm1)$ of $\sim 2\pi\times 149\,\text{Hz}$ leads to a dynamic evolution of the phase $\phi = \varphi_0 - (\varphi_{+1}+\varphi_{-1})/2$, with the individual phases $\varphi_j$ of the magnetic substates.
After a variable time, the longitudinal confining potential is switched off. 
After $6\,$ms of expansion we apply a radiofrequency (rf) pulse resonant with the transition $(1,0)\leftrightarrow(1,\pm1)$.
This can be modeled as a spin rotation described by the Hamiltonian $\hat{H}\ind{rf} = \Omega\ind{rf}\left[i\hat{a}^\dagger_{0}(\hat{a}_{+1} - \hat{a}_{-1}) + {\rm h.c.}\right]/\sqrt{2}$ with the resonant Rabi frequency  $\Omega\ind{rf} = 2\pi\times6.25$\,kHz.
Applying a $\pi/2$-pulse of duration $40\,\mu$s the observable $\hat F(\phi)=\left[(\hat{a}^\dagger_{+1}+\hat{a}^\dagger_{-1})e^{i\phi}\hat{a}_{0}+\text{h.c.}\right]/\sqrt{2}$ is directly mapped to the measurable population difference $\hat N^-=\hat N_{+1}-\hat N_{-1}$.
With the population $N_0$ after this spin rotation we access the observable ($\hat{N}_{+1}+\hat{N}_{-1}+\hat{a}^\dagger_{+1}\hat{a}_{-1}+\hat{a}^\dagger_{-1}\hat{a}_{+1})/2$ before the rotation,  which we use to estimate corrections to the commutator between $\hat F(0)$ and $\hat F(\pi/2)$ (see below).\\
After expansion we apply a Stern-Gerlach magnetic field gradient pulse followed by a short time of flight of $\approx 1\,$ms to separate the three spin components spatially and thus enable a state-selective read-out. 
We use destructive absorption imaging in the strong saturation regime by applying a $\tau = 15\,\mu$s resonant light pulse.
\begin{figure}[!t]
	\centering
	\includegraphics[width = 0.8\columnwidth]{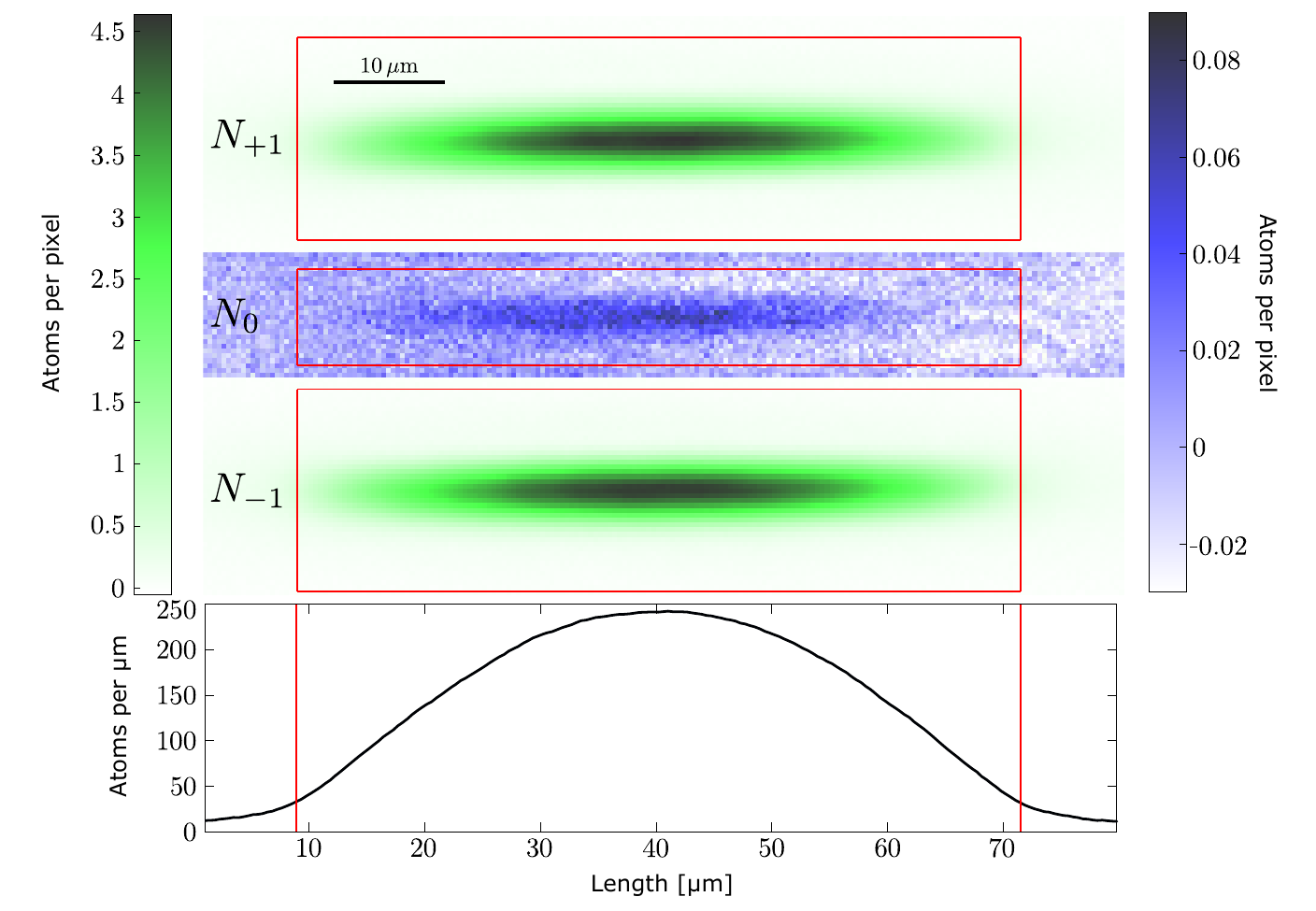}
	\caption{\textbf{Absorption imaging and analysis regions.} After an expansion time of 6\,ms we apply a resonant rf $\pi/2$-pulse, and subsequently a Stern-Gerlach gradient pulse, which allows for spatially resolved read-out of the populations of $(1,\pm 1)$ and $(1,0)$. We show the average of the absorption signal over $1247$ experimental realizations after 150\,ms of spin mixing time. The red lines indicate the evaluation regions. Two different color scales (left scale for $(1,\pm1)$, right for $(1,0)$) are chosen. The population in $(1,\pm1)$ is significantly larger than the one in $(1,0)$, which sums to $N_0\approx 25$ atoms in total. 
		The lowest panel shows the column integrated signal over all spin states.}
	\label{MeanAbsorption150}
\end{figure}
\subsection*{Classical correlations in absorption imaging}
\begin{figure}[t!]
	\linespread{1}
	\centering
	\includegraphics[width = 0.5\columnwidth]{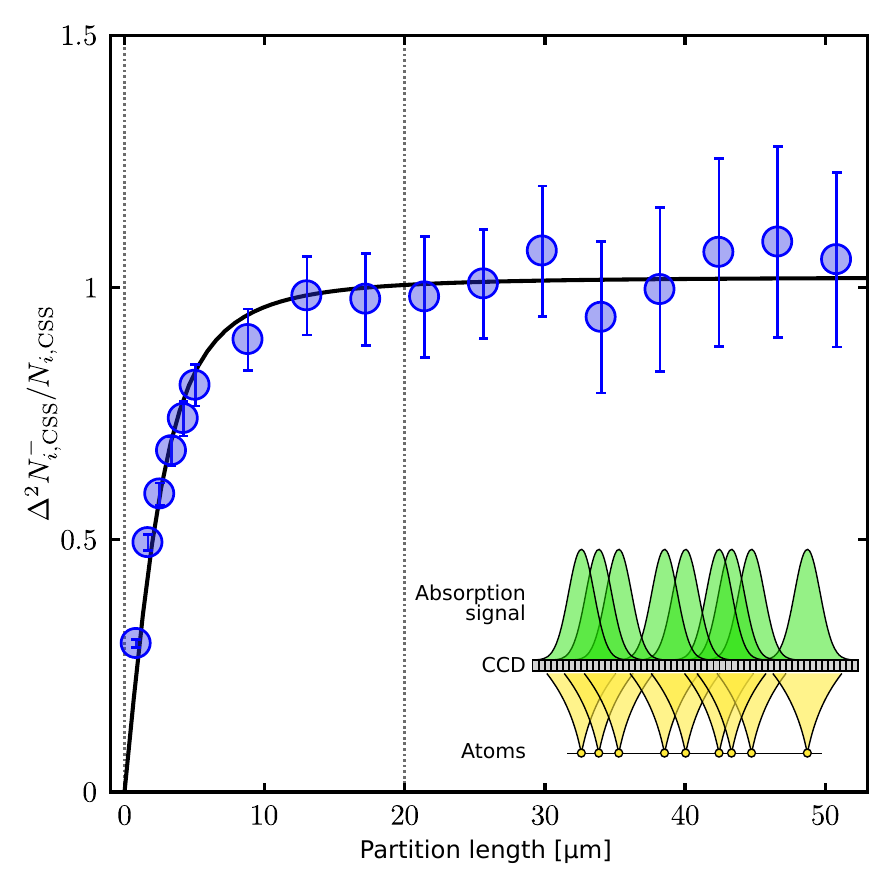}
	\caption{\textbf{Classical correlations in absorption imaging.} After $\pi/2$ rf-rotation of the state $(1,0)$ we analyze the fluctuations $\Delta^2N^{-}_{i,\text{CSS}}/N_{i,\text{CSS}}$ for partitions of equal length of the absorption signal. For small partitions we find reduced fluctuations compared to the expectation of binomial statistics ($\Delta^2N^{-}_\text{bin}/N = 4p(1-p)=1$ with $p = N_{+1}/N=1/2$). One contribution to this effect is the blurring of the atomic signal over a finite region on the detecting CCD camera (see inset). The line is a fit to the scaled suppression factor $\zeta(x)=\sqrt{bx^2/(1+bx^2)}$ with the partition length $x$. Using a Monte-Carlo simulation with a Gaussian spread function we find that this corresponds to an rms-radius $\sim 1.2\,\mu$m of the image of every atom on the detector. For the minimal partition length of 20\,$\mu$m chosen in the steering analysis classical correlations are negligible.}
	\label{Blurring}
\end{figure}

As described in the main text, classical correlations of the absorption signal are dominant below a certain length scale. 
To characterize this effect we analyze partitions of the absorption signal of a fully separable coherent spin state (CSS). We prepare this state by rf rotation ($\pi/2$ pulse) of our initial state $(1,0)$, resulting in an equal superposition of $(1,+1)$ and $(1,-1)$ with negligible $(1,0)$ component.
Before the rf pulse we clean spurious population in $(1,\pm1)$ by applying a strong Stern-Gerlach pulse and microwave $\pi$-pulses of $(1,\pm1)$ to $(2,\pm1)$. 
When analyzing parts of the cloud by partitioning the absorption signal we observe reduced fluctuations of the detected particle number difference, i.e. $\Delta^2 N_{i,\text{CSS}}^-/N_{i,\text{CSS}}$ compared to the expected binomial statistics (see Figure~\ref{Blurring}), resulting from classical correlations of neighboring spatial regions.
This can arise due to a blurring of the absorption signal during the imaging process (also described in \cite{Greiner2005,Esteve2006}) e.g. caused by finite optical resolution or the lateral random walk of the atoms during photon scattering.
Thereby a single atom is imaged to a finite region on the detecting CCD camera.
We simulate this effect using a Monte-Carlo approach by dicing atomic positions and generating the corresponding absorption signal with a characteristic spatial spread of each atom (see inset Figure~\ref{Blurring}). We partition the resulting signal spatially and find that the reduction of the local fluctuations is well described by a suppression factor $\zeta(x) = \sqrt{bx^2/(1+bx^2)}$, where $x$ is the length of the partition.
The resulting reduction of $\Delta^2 N_{i,\text{CSS}}^-/N_{i,\text{CSS}}$ is $50$\% for $x_{1/2} = 1/\sqrt{3b}$.
For a Gaussian spatial spread function this corresponds to approximately twice the rms radius $s_\text{rms}$.
A fit of $\zeta(x)$ to the experimental data gives $s\ind{rms} \approx 1.2\,\mu$m (see Figure~\ref{Blurring}). 
Our optical resolution including the push of the atoms through the focal plane during the imaging pulse~\cite{Muessel2013} amounts to $s_\text{opt}\sim 0.5\,\mu$m. An upper limit for the lateral jiggle due to random photon recoils is given by~\cite{Joffe1993} $s_\text{jig} = v_\text{rec}\sqrt{\Gamma_\text{sc}}\tau^{3/2}/3\sim 0.5\,\mu$m. Here, $\Gamma_\text{sc} \approx 2\pi \times 6\,\text{MHz}/2$ is the scattering rate, $v_\text{rec}\approx6\,\text{mm}/\text{s}$ the recoil velocity and $\tau=15\,\mu$s the duration of the imaging pulse. As a combined effect, we get $\sqrt{s_\text{opt}^2 + s_\text{jig}^2} \sim 0.7\,\mu$m, suggesting that these are the main causes for the classical correlations.
To keep the influence of classical correlation effects negligible, we restrict the partition size in the analysis of EPR steering to $>20\,\mu$m which is well in the saturation regime of the obtained $\zeta(x)$ (see Figure~\ref{Blurring}).

\subsection*{Commutation relations and Larmor phase}

The full expression for the commutator of the spin observables $\hat{F}(\phi)$, appearing in the EPR steering bound, reads
\begin{equation}
 \left[\hat{F}(\phi_1),\hat{F}(\phi_2)\right] = i \sin(\phi_2-\phi_1) \left[2\hat N_0 - (\hat a_1^\dagger +\hat a_{-1}^\dagger)(\hat a_1 +\hat a_{-1})\right] \,.
\end{equation}
Thus, for $\phi_2-\phi_1=\pi/2$, neglecting the correction terms $\mathcal{O} ( N_1+ N_{-1})$, one obtains $[\hat{F}(\phi_1),\hat{F}(\phi_2)]=2i\hat N_0$. 
We stress that the expectation value of the correction term can be directly measured.
For spin mixing times smaller than $150\,$ms we find the relative deviation of the commutator from $2N_0$ to be below $0.25$\% (see Fig.~\ref{MeanAbsorption150}).

We note that the Larmor phase $\phi_\text{L} = (\varphi_{+1}-\varphi_{-1})/2$ between the components $(1,+1)$ and $(1,-1)$ cannot be controlled experimentally. Thus, strictly speaking, in every repetition of the experiment, the observable 
\begin{equation}
 \hat F(\phi,\phi_\text{L}) = \frac{1}{\sqrt{2}} \left[(e^{-i\phi_\text{L}}\hat{a}^\dagger_{1}+e^{i\phi_\text{L}}\hat{a}^\dagger_{-1})e^{i\phi}\hat{a}_{0}+\text{h.c.}\right]
\end{equation}
is measured with random $\phi_\text{L}$. 
However, this does not alter our conclusions about the steering bound derived above, since the product of the inference variances averaged over all Larmor phases is still bounded by the commutator at equal $\phi_\text{L}$. We denote this bound by $b=|\langle[\hat F(\phi,\phi_\text{L}),\hat F(\phi+\pi/2,\phi_\text{L})]\rangle|^2/4$. This can be seen by estimating
\begin{equation}
\begin{split}
 \frac{1}{N_\text{L}} \sum_i\Delta^2\hat F(\phi,\phi_{\text{L},i}) \frac{1}{N_\text{L}} \sum_j\Delta^2\hat F(\phi+\pi/2,\phi_{\text{L},j}) &\geq \frac{1}{N_\text{L}^2} \sum_i\Delta^2\hat F(\phi,\phi_{\text{L},i}) \sum_j \frac{b}{\Delta^2\hat F(\phi,\phi_{\text{L},j})} \\
 &=\frac{b}{N_\text{L}^2}\sum_{i,j} \eta_{ij} \\
 &=\frac{b}{N_\text{L}^2}\left[ N_\text{L} + \sum_{i<j}\left( \eta_{ij} + \frac{1}{\eta_{ij}}   \right) \right] \\
 & \geq \frac{b}{N_\text{L}^2} \left[N_\text{L} + 2\,\frac{N_\text{L}(N_\text{L}-1)}{2}  \right] = b\,,
 \end{split}
\end{equation}
where the $\phi_{\text{L},i}$ form a set of $N_\text{L}$ Larmor phases, and $\eta_{ij} = \Delta^2\hat F(\phi,\phi_{\text{L},i})/\Delta^2\hat F(\phi,\phi_{\text{L},j}) $. Here, we used that $\eta_{ij}+1/\eta_{ij}\geq 2$ for $\eta_{ij}>0$. Formally, the summation over the discrete set of Larmor phases can be converted into an integral. Experimentally, each Larmor phase is sampled with finite statistics, and the only assumption we make is that it is not biased, i.e., that all phases are realized with the same probability. We note that, for states obtained from time evolution under the spin mixing Hamiltonian, the inference variances are in fact independent of the Larmor phase.

\subsection*{Witness for genuine multipartite entanglement}

Here, we outline the derivation of the witness for genuine $m$-partite entanglement used in the main text. EPR steering of subsystem A by B, indicated by the steering product $S_{\rm A|B}<1$, implies that the subsystems are entangled. We seek to derive a corresponding bound on $S_{\rm A|B}$ the violation of which indicates genuine $m$-partite entanglement. For this, we subdivide B into $m-1$ parts of equal size (atom number) to obtain an $m$-partite system. Genuine $m$-partite entanglement with respect to such a partitioning means that the state of the system is not biseparable, i.e. it is not separable with respect to any division of the $m$ parties into two groups X and $\rm \overline X$, nor is it a mixture of biseparable states \cite{Teh2014}. 

To derive the bound on $S_{\rm A|B}$ we recall that the inference variances (3) involve operators of the form
\begin{equation}
  \hat u=\sum_i g_i \hat F_i(0)\,, \qquad \hat v=\sum_i h_i \hat F_i(\pi/2)\,,
\end{equation}
where $g_i$, $h_i$ are real numbers, and $\hat F_i(\phi)$ acts on subsystem $i$.
With this, we derive that for a biseparable state the inequality
\begin{equation}
 N_\text{A}\sqrt{S_{\rm A|B}} = \Delta u\, \Delta v \geq \min_\alpha \left[ \left|\sum_{i\in {\rm X}_\alpha}g_i h_i N_i\right| + \left|\sum_{i\in {\rm\overline X}_\alpha}g_i h_i N_i\right| \right]
\end{equation}
holds (see Supplementary Text for details). Thus, if this inequality is violated, the state must be genuinely $m$-partite entangled. Here, $N_i$ is the mean atom number in subsystem $i$. The index $\alpha$ labels all possible ways to divide the $m$ subsystems into two groups.

Experimentally (cf.\ Fig.~\ref{Multipartite}) we choose $g_1=h_1=1$ and $g_{i>1}=g$, $h_{i>1}=h$, where $g$ and $h$ are chosen such that the inference variances $\Delta u$ and $\Delta v$ are minimal. The $m$ subsystems consist of A with $N_1=N_\text{A}=\eta_\text{A} N$ atoms and the sizes (atom numbers) of the subsystems $N_{i>1}=N(1-\eta_\text{A})/(m-1)$ are all chosen to be equal. In this case, the inequality (10) simplifies to
\begin{equation}
  \Delta u \,\Delta v \geq \min_{k=1\ldots m-1} \left[ \left| \eta_\text{A} N+ (k-1)g h \frac{N(1-\eta_\text{A})}{m-1}\right| + (m-k)|g h| \frac{N(1-\eta_\text{A})}{m-1} \right]
\end{equation}
For the values of $g=g(0)>0$ and $h=g(\pi/2)<0$ the minimum is always attained for $k=m-1$, which gives 
\begin{equation}
  \sqrt{S_{\rm A|B}} = \frac{\Delta u \,\Delta v }{N_\text{A}}\geq 1 + \frac{m-3}{m-1}\,g(0)\, g(\pi/2)\,\frac{1-\eta_\text{A}}{\eta_\text{A}} 
\end{equation}
which can be rearranged to give eq.~(5).

\vspace*{1cm}

\clearpage

\section*{Supplementary Text}
\subsection*{Observables and SU(3) algebra}

In the main text we define the observables 
\begin{equation}
  \hat{F}(\phi)= \frac{1}{\sqrt{2}}\left[(\hat{a}^\dagger_{+1}+\hat{a}^\dagger_{-1})\mathrm e^{i\phi}\hat{a}_{0} + \text{h.c.}\right]\,,
\end{equation}
where the phase $\phi_0$ has been set to zero.
These observables can be expressed in terms of the SU(3) generators consisting of the three components $\hat S_\alpha$ of the spin vector and elements $\hat Q_{\alpha\beta}$ of the quadrupole tensor, where $\alpha,\beta\in\{x,y,z\}$ \cite{Hamley2012}, as $\hat{F}(\phi) = \cos(\phi)\hat S_x  + \sin(\phi)\hat Q_{yz}$. 
The radiofrequency pulse can be viewed as a rotation in the SU(2)-subspace generated by $\left\{\hat S_x,\hat S_y,\hat S_z\right\}\,$.
In general, one has $\hat{H}_{\rm{rf}}=\Omega \left[\cos(\phi\ind{L})\hat S_y + \sin(\phi\ind{L})\hat S_x \right]$.
With this, the measured observable becomes $\hat F(\phi,\phi\ind{L}) = \cos(\phi)\hat S_\perp + \sin(\phi)\hat Q_\perp$, with $\hat S_\perp =  \cos(\phi\ind{L})\hat S_y - \sin(\phi\ind{L})\hat S_x$ and $\hat Q_\perp =  \cos(\phi\ind{L})\hat Q_{xz} + \sin(\phi\ind{L})\hat Q_{yz}$.

\subsection*{Theoretical modeling of entanglement generation and detection}

\paragraph*{Spin mixing dynamics:}

Under the assumption that the atoms in each hyperfine component populate only a single spatial mode, the Hamiltonian describing the internal-state dynamics is given by \cite{Gerving2012} (see also Methods section)
\begin{equation}
\label{eq:H}
\hat{H}\ind{SM}=\lambda \left(2\hat N_0-1\right)\left(\hat N_{+1}+\hat N_{-1}\right) + 2 \lambda\left(\hat a_0^\dagger \hat a_0^\dagger \hat a_{+1} \hat a_{-1} + \hat a_{+1}^\dagger \hat a_{-1}^\dagger \hat a_0 \hat a_0 \right) + q\left(\hat N_{+1}+\hat N_{-1}\right)\,,
\end{equation}
where $\lambda$ is the collisional interaction coefficient, which is negative for the $F=1$ state of $^{87}$Rb, and the experimentally adjustable detuning $q$. 

This Hamiltonian conserves both the total atom number, $ N= N_0+ N_{+1}+ N_{-1}$, and the magnetization, or population difference, $ M= N_{+1}- N_{-1}$. Since the initial state is $\ket{N_{-1},N_0,N_{+1}}=\ket{0,N,0}$, with $M=0$, only states with $N_{+1}=N_{-1}\equiv n$ are populated. Thus the states of the relevant Hilbert space can be labeled by the number of pairs $n$ ($n=0\ldots N/2$). Writing the Hamiltonian \eqref{eq:H} in this basis results in a tridiagonal matrix and the Schr\"odinger equation can be straightforwardly integrated numerically.

For short evolution times and large $N$ the population of the $m_{\rm F}=0$ state is much larger than that of the side modes $m_{\rm F}=\pm 1$. In this limit the operator $\hat a_0$ can be replaced by a c-number $\hat a_0\rightarrow \sqrt{N} \mathrm e^{i\varphi_0}$. Under this approximation $q$ can be tuned to cancel the elastic collision term, i.e.\ the first term in eq.~\eqref{eq:H}, leading to $\hat{H}\ind{SCC}=2\lambda N(\mathrm e^{-2i\varphi_0}\hat a_{+1} \hat a_{-1}+\mathrm e^{2i\varphi_0}\hat a_{+1}^\dagger \hat a_{-1}^\dagger)$ reminiscent of the undepleted-pump approximation for parametric down conversion in quantum optics. The evolution under this Hamiltonian can be calculated analytically, giving
\begin{equation}
\ket{\psi(t)} = \frac{1}{\cosh(r)}\sum_n (-i \mathrm e^{-2i\varphi_0} x)^n\ket{n,N-2n,n} \,,
\end{equation}
with $x=\tanh(r)$ and the squeezing parameter $r=2N\lambda t$. In the following we choose $\varphi_0 =0$ and define $c_n=(-ix)^n/\cosh(r)$.

\begin{figure}[!t]
	\linespread{1}
  \centering
 \includegraphics[width=0.65\columnwidth]{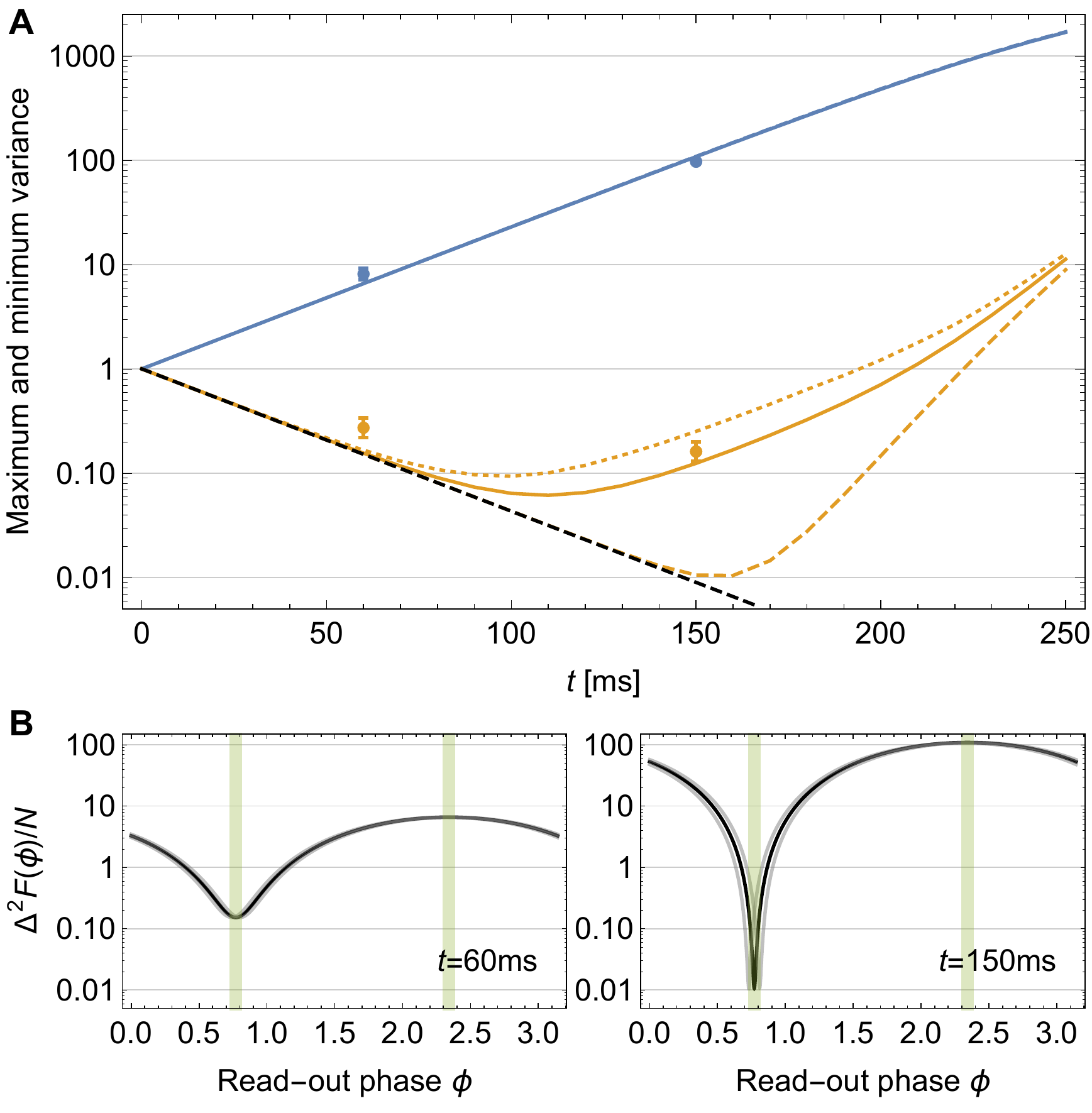}
 \caption{\textbf{Squeezing and anti-squeezing of the global spin.} (\textbf{A}) Minimum (orange) and maximum (blue) variance as a function of spin mixing time. Here, $2N\lambda/2\pi=-2.5\,$Hz, $q/2\pi=2.4\,$Hz, $N=10^4$. Black dashed: ideal case in undepleted-pump approximation. Orange dashed: ideal case including pump depletion. Solid: including fluctuations in $q/2\pi$ by $\pm 0.2\,$Hz. Dotted: assuming that the point of minimal uncertainty lies exactly between two measured settings of $\phi$ (worst case), i.e., the variance was evaluated half a scan step away from the position of the minimum (maximum). The chosen minimum experimental scan step was $\Delta\phi=0.03\pi$. One observes that at long spin mixing times the maximum achievable squeezing is limited by the fluctuations in $q$ since the $\phi$ interval in which squeezing is present, becomes small (see panel (\textbf{B})). For the anti-squeezed quadrature (blue) all three lines (solid, dashed, and dotted) coincide, showing that this variance is insensitive with respect to the discussed imperfections. Therefore the value of $\lambda$ is adjusted to match the rate of the exponential increase of the maximal variance. (\textbf{B}) Dependence of the variance on the tomography angle. The gray lines indicate the effect of a drift of $q/2\pi$ by $\pm 0.2\,$Hz. The green shaded regions show the size of the minimum angle scan step used in the experiment.}
 \label{fig:squeezing}
\end{figure}

Within the undepleted-pump approximation it is straightforward to obtain analytical expressions for the side-mode population and the variance of $\hat{F}(\phi)$, which read
\begin{align}
 \langle \hat N_{+1} \rangle = \langle \hat N_{-1}\rangle &= \sum_n n|c_n|^2 = \sinh^2 (r) \,, \\ 
 \Delta^2 \hat{F}(\phi) = \langle \hat{F}^2(\phi) \rangle &= N\,[\cosh(2r)-\sinh (2r) \sin(2\phi)] \, ,
\end{align}
where we used that $\langle\hat{F}(\phi)\rangle = 0$.
Thus, the minimum variance is $\Delta^2 \hat{F}(\pi/4)/N=\exp(-2r)$ (squeezed quadrature) and the maximal variance is $\Delta^2 \hat{F}(3\pi/4)/N=\exp(2r)$ (anti-squeezed quadrature). In the main text we absorbed $\pi/4$ into the offset phase $\phi_0$ and defined the cases of maximal and minimum variance as $\hat{F}(0)$ and $\hat{F}(\pi/2)$. Experimentally $\phi_0$ is an arbitrary but fixed phase offset. 
In Fig.~\ref{fig:squeezing}A, we compare the result for the minimum variance in undepleted-pump approximation (black dashed line) to numerical integration of the Schr\"odinger equation (yellow dashed line). This shows agreement for evolution times $t \lesssim 150\,$ms for our experimental parameters. For longer spin mixing times the depletion of the $m_{\rm F}=0$ population becomes relevant and the minimum variance deviates from the initial exponential decay and increases again.

\paragraph*{Division into subsystems:}

Next, we consider splitting the cloud into two halves. If the two parts A and B of the cloud are of equal size, we can assume that any atom is detected in A or B with equal probability and independently of the other atoms. We define mode operators $\hat a_{i,\text{A}}$ and $\hat a_{i,\text{B}}$ which destroy a particle in subsystem A and B, respectively, such that $\hat a_i= (\hat a_{i,\text{A}} + \hat a_{i,\text{B}})/\sqrt{2}$, where $i\in\{-1,0,+1\}$. This is equivalent to a beamsplitter operation in optics \cite{Armstrong2012, Armstrong2015}, where one defines two orthogonal input modes $\hat a_i$ and $\hat a_{i,\rm aux}=(\hat a_{i,\text{A}} - \hat a_{i,\text{B}})/\sqrt{2}$, the latter being in a vacuum state.
Mathematically, this can be viewed as extending the Hilbert space by adding an auxiliary mode for each internal state. As these modes are never populated the number of basis states spanning the relevant Fock space is unchanged: $\ket{N_{-1},N_0,N_{+1},N_{-1,\text{aux}},N_{0,\text{aux}},N_{+1,\text{aux}}}=\ket{n,N-2n,n,0,0,0}$. The output modes of the beam splitter are thus $\hat a_{i,\text{A}}=(\hat a_i+\hat a_{i,\text{aux}})/\sqrt{2}$, $\hat a_{i,\text{B}}=(\hat a_i-\hat a_{i,\text{aux}})/\sqrt{2}$. We can now express $\hat F_\text{A}(\phi)$ and $\hat F_\text{B}(\phi)$ in terms of the two input modes of the beamsplitter,
\begin{equation}
\begin{aligned}
 \hat F_{\text{A}}(\phi) &= \frac{1}{\sqrt{2}}\left[\mathrm e^{-i\phi}\hat a_{0,\text{A}}^\dagger(\hat a_{+1,\text{A}}+\hat a_{-1,\text{A}}) + {\rm h.c.}\right] \\
 &= \frac{1}{2\sqrt{2}}\left[\mathrm e^{-i\phi} (\hat a_0^\dagger +\hat a_{0,\text{aux}}^\dagger )(\hat a_{+1}+\hat a_{+1,\text{aux}} + \hat a_{-1} + \hat a_{-1,\text{aux}}) + {\rm h.c.}\right] \,,
\end{aligned}
\end{equation}
and similarly for $\hat F_{\text{B}}(\phi)$. This is easily generalized to the case of asymmetric splitting $\hat a_i= \sqrt{\eta_\text{A}} \,\hat a_{i,\text{A}} + \sqrt{\eta_\text{B}} \,\hat a_{i,\text{B}}$ (with $\eta_\text{A} + \eta_\text{B}=1$). In the case of splitting the cloud into $3$ parts a three-port beam splitter picture can be used where, in addition to $\hat a_i= \sqrt{\eta_\text{A}}\, \hat a_{i,\text{A}} + \sqrt{\eta_\text{B}}\, \hat a_{i,\text{B}} + \sqrt{\eta_\text{C}}\, \hat a_{i,\text{C}}$, we define two auxiliary modes, such that all input modes are orthogonal to each other, and invert this linear transformation to obtain the outputs in terms of the inputs. 

With this, we can calculate the relevant observables in terms of the coefficients $c_n$ (here for the case of symmetric splitting into two parts). For the variance we obtain
\begin{equation}
\begin{aligned}
 \langle \hat F_\text{A}^2(\phi) \rangle &= \langle \hat F_\text{B}^2(\phi) \rangle \\ 
  &= \sum_n |c_n|^2 \frac{N_0(n+1)+n}{2} +\left( \mathrm e^{2i\phi} c_n c^*_{n+1} \frac{\sqrt{N_0(N_0-1)}(n+1)}{4} + \rm{c.c.}\right) \\
  &\approx N \sum_n |c_n|^2 \frac{n+1}{2} + \left(\mathrm e^{2i\phi} c_n c^*_{n+1} \frac{n+1}{4} + \rm{c.c.} \right),
\end{aligned}
\end{equation}
where c.c. denotes the complex conjugate, and for the covariance
\begin{equation}
\begin{aligned}
 \langle \hat F_\text{A}(\phi)\hat F_\text{B}(\phi) \rangle &= \langle \hat F_\text{B}(\phi)\hat F_\text{A}(\phi) \rangle \\ 
  &= \sum_n |c_n|^2 \frac{N_0 n}{2} + \left(\mathrm e^{2i\phi} c_n c^*_{n+1} \frac{\sqrt{N_0(N_0-1)}(n+1)}{4} + \rm{c.c.}\right) \\
  &\approx N \sum_n |c_n|^2 \frac{n}{2} + \left(\mathrm e^{2i\phi} c_n c^*_{n+1} \frac{n+1}{4} + \rm{c.c.}\right)\,.
\end{aligned}
\end{equation}
The last line in each equation uses the undepleted-pump approximation (only keeping terms proportional to $N$ ($N_0=N-2n)$). In this limit  we can use the analytical expression for $c_n$ to obtain
\begin{equation}
 \frac{\langle \hat F_\text{A}^2(\phi)\rangle}{N/2} =\frac{1}{2}\left[1+\cosh(2r)-\sinh(2r)\sin(2\phi)\right]
 \label{eq:varA}
\end{equation}
and 
\begin{equation}
 \frac{\langle \hat F_\text{A}(\phi)\hat F_\text{B}(\phi)\rangle}{N/2} = \frac{1}{2}\left[-1+\cosh(2r)-\sinh(2r)\sin(2\phi)\right] \, .
 \label{eq:covar}
\end{equation}
Generalizing the above calculation to the case of splitting the cloud into unequal parts yields, in undepleted-pump approximation,
\begin{align}
 \frac{\langle \hat F_\text{A}^2(\phi)\rangle}{\eta_\text{A} N} &= 1+\eta_A\left[-1+\cosh(2r)-\sinh(2r)\sin(2\phi)\right] \,,\\
 \frac{\langle \hat F_\text{B}^2(\phi)\rangle}{\eta_\text{B} N} &= 1+\eta_B\left[-1+\cosh(2r)-\sinh(2r)\sin(2\phi)\right] \,,\\
 \frac{\langle \hat F_\text{A}(\phi)\hat F_\text{B}(\phi)\rangle}{\eta_\text{A} \eta_\text{B} N} &=-1+\cosh(2r)-\sinh(2r)\sin(2\phi) \, .
\end{align} 

From this, we calculate the inferred variances used to demonstrate EPR steering which are defined as $\Delta^2  F_{\text{A}|\text{B}}(\phi) = \text{min}_{g(\phi)}[\Delta^2(\hat F_\text{A}(\phi)-g(\phi)\hat F_\text{B}(\phi))]$. The minimization gives $g(\phi)=\langle \hat F_\text{A}(\phi) \hat F_\text{B}(\phi) \rangle/\langle \hat F_\text{B}^2(\phi) \rangle$ and thus $\Delta^2  F_{\text{A}|\text{B}}(\phi) =\langle \hat F_\text{A}^2(\phi) \rangle - \langle \hat F_\text{A}(\phi) \hat F_\text{B}(\phi) \rangle^2/\langle \hat F_\text{B}^2(\phi) \rangle$. The solid black lines in Fig.~\ref{Bipartite}A and B of the main text are obtained by integrating the full Schr\"odinger equation for $N=10^4$ atoms and using the resulting coefficients $c_n$ as input to equations \eqref{eq:varA} and \eqref{eq:covar} to calculate $\Delta^2  F_{\text{A}}(\phi)$ and $\Delta^2  F_{\text{A}|\text{B}}(\phi)$, with the experimental parameters given in the caption of Fig.~\ref{fig:squeezing}.

\begin{figure}
  \centering
 \includegraphics[width=0.9\columnwidth]{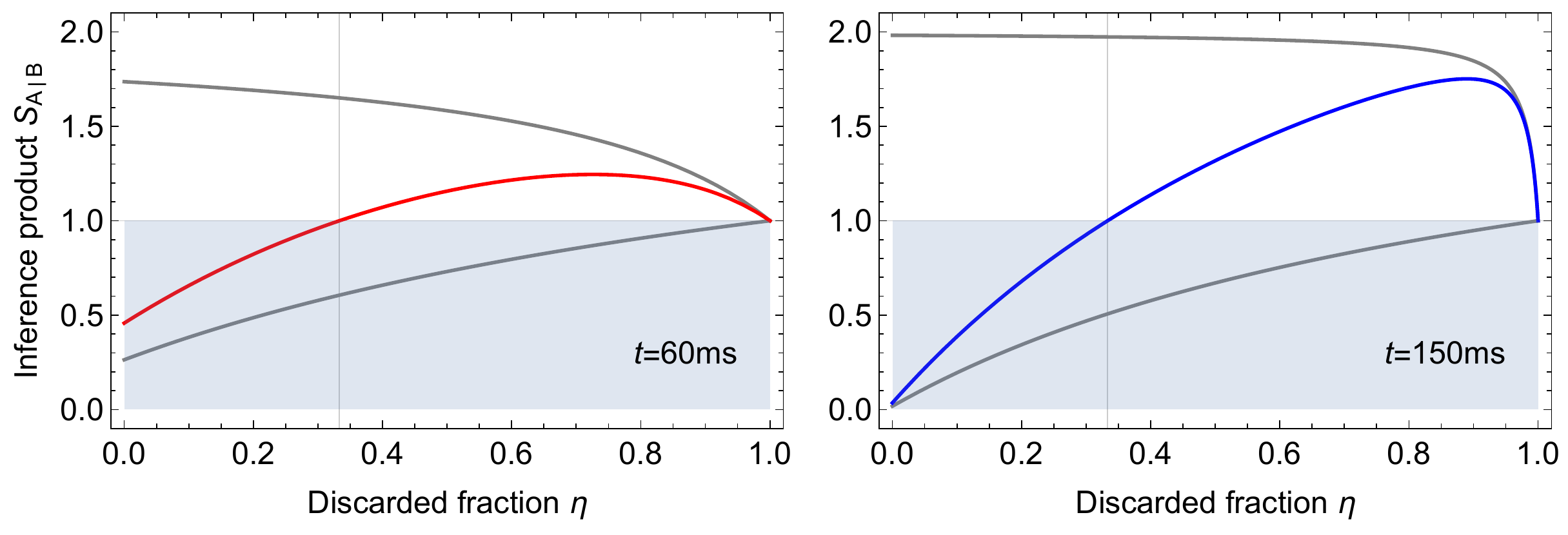}
 \caption{\textbf{Discarding part of the cloud.} We show the inference variances of the squeezed ($\phi=\pi/4$) and anti-squeezed ($\phi=3\pi/4$) quadratures (gray) and their product (red and blue, respectively) at two different spin mixing times. The discarded fraction $\eta=1-\eta_A-\eta_B$ is varied. The remaining cloud is divided into equal parts, $\eta_A=\eta_B$, to calculate the steering product $S_{\rm A|B}$.}
 \label{fig:discard_part}
\end{figure}

In the case of splitting the cloud into three parts A, B, and C, discarding C and steering A with B one obtains exactly the same expressions, with the only difference that the constraint $\eta_\text{A} + \eta_\text{B}=1$ is relaxed to $\eta_\text{A} + \eta_\text{B}\leq 1$.

We illustrate the inference variance for the squeezed and anti-squeezed quadratures and their product in Fig.~\ref{fig:discard_part}. Here, we discard a fraction $\eta$ of the signal and divide the remaining cloud into equal parts A and B, with $\eta_A=\eta_B = (1-\eta)/2$. We show the inference variances for steering A with B as a function of $\eta$, as done in Fig.~\ref{Bipartite}C in the main text. This illustrates that steering is not possible if more than one third of the signal is discarded. Note that this is consistent with the monogamy of steering, which requires that for a tripartite system, consisting of parts A, B, and C, if A is steerable by B, A cannot be steerable by C. Formulated in terms of steering products, this condition reads $S_{ \text{A}|\text{B}} S_{ \text{A}|\text{C}}\geq 1$ \cite{Reid2013}.

At this point a comment on the validity of the beam splitter picture is in order. Experimentally, we observe that the fluctuations of the total atom number between the two halves of the cloud are suppressed as compared to binomial statistics expected from a linear beam splitter. To gauge possible effects of this on the observed EPR steering, we consider the extreme case that the cloud is always split into two halves containing exactly $N/2$ atoms. This is achieved by viewing the atoms as distinguishable particles (spins) subject to all-to-all interactions.

Since the Hamiltonian $\hat{H}\ind{SM}$ \eqref{eq:H} conserves the total particle number, it can be expressed in terms of collective-spin operators using
\begin{equation}
 \hat \sigma_{\alpha\beta} =\ket{\alpha}\bra{\beta} \,,
\end{equation}
\begin{equation}
 \hat a_\alpha^\dagger \hat a_{\beta}\to\hat S_{\alpha\beta} = \sum_{i=1}^N \hat \sigma_{\alpha\beta}^{(i)}\,.
\end{equation}
The relevant Hilbert space is still spanned by states $\ket{N_{+1},N_0,N_{-1}}$. When expressed in terms of the canonical basis sates of the Hilbert space of $N$ distinguishable spin $1$ particles, these states form the subspace of fully symmetrized states
\begin{equation}
 \ket{N_{+1},N_0,N_{-1}} = \sqrt{\frac{N_{+1}!N_0!N_{-1}!}{N!}} \hat{S}\big[\ket{1,1,\ldots, 0,0, \ldots, -1,-1,\ldots}\big]\, ,
\end{equation}
where $\hat{S}[\cdot]$ denotes the symmetrization operator, i.e.\ the sum over all states with $N_{+1}$ spins in state $m_{\rm F}=1$, $N_0$ in state $m_{\rm F}=0$, and $N_{-1}$ in state $m_{\rm F}=-1$. The multinomial coefficient ensures normalization. 
The spin-changing collision Hamiltonian thus translates to
\begin{equation}
 \hat{H}\ind{SCC} = 2\lambda (\hat a_0^\dagger \hat a_0^\dagger \hat a_{+1} \hat a_{-1} + \hat a_{+1}^\dagger \hat a_{-1}^\dagger \hat a_0 \hat a_0 ) \to 2\lambda(\hat S_{01}\hat S_{0-1}+\hat S_{10}\hat S_{-10})\,.
\end{equation}

Within this spin picture, operators acting on a subsystem of $N_\text{A}$ spins are now defined by summing only over $N_\text{A}$ single-spin operators $\hat S_{\alpha\beta,\text{A}} = \sum_{i=1}^{N_\text{A}} \hat \sigma_{\alpha\beta}^{(i)}$. Given a state in the basis $\ket{N_{+1},N_0,N_{-1}}$, we calculate the expectation value of, e.g., $\hat F_\text{A}(\phi)$ defined as 
\begin{equation}
 \hat F_\text{A}(\phi) =  \frac{1}{\sqrt{2}} \left[\mathrm e^{-i\phi}(\hat S_{10,\text{A}}+\hat S_{-10,\text{A}}) +\mathrm e^{i\phi}(\hat S_{01,\text{A}}+\hat S_{0-1,\text{A}}) \right] \,. 
\end{equation}
For this, the basis states have to be decomposed into a basis of product states between subsystems A and B, which gives
\begin{equation}
 \ket{N_{+1},N_0,N_{-1}} = \sqrt{\frac{N_0!}{\binom{N}{N_\text{A}}}} \sum_{k,p} \sqrt{\frac{\binom{n}{k}\binom{n}{p}}{N_{0,\text{A}}!N_{0,\text{B}}!}}\ket{k,N_{0,\text{A}},p}\otimes\ket{n-k,N_{0,\text{B}},n-p} \,.
 \label{eq:spin_split}
\end{equation}
With this, we can evaluate the expectation values for all quantities relevant for the EPR steering bound assuming that the ensemble is split into subsystems of precisely $N_\text{A}$ and $N-N_\text{A}$ atoms with zero fluctuations. For $N_\text{A},N_\text{B},N\gg 1$ the relative fluctuations $\Delta N_\text{A}/N_\text{A}\ll 1$ and thus the atom number fluctuations are irrelevant and the two pictures equivalent. We have confirmed numerically that for the parameters of our experiment ($N\approx 10^4$) the spin picture leads to the same results as the beamsplitter description.

This shows that for the entanglement distribution, it is not crucial that the atomic cloud expands in a fully self-similar fashion. The observed entanglement is due to the fact that the pair creation process is permutation invariant and populates highly particle-entangled collective states. This is what leads to the robustness of the generated entanglement with respect to the expansion in the waveguide. The fact that entanglement due to bosonic symmetrization of indistinguishable particles can be made accessible by elementary operations (here: self-similar expansion) has been the subject of a long debate (see e.g. \cite{Killoran2014} and references therein). Thus, our experiment demonstrates that entanglement of indistinguishable particles in a single spatial mode is, in the sense of the LOCC paradigm, as useful as entanglement between distinguishable particles.

\paragraph*{Fluctuations of the detuning $\boldsymbol{q}$:}

The main experimental imperfection that limits the achievable degree of steering is slow drifting of magnitude of the ac-Zeeman shift, which leads to drifts of the fringe position. We can account for this by using different values of $q$ in our simulation and calculate weighted averages of the outcomes. Moreover, at long spin mixing times the range of phases $\phi$ where the variance is squeezed becomes small such that a coarse scan of the angle might miss the minimum. These effects are illustrated in Fig.~\ref{fig:squeezing}.

\subsection*{Bounds for genuine $\boldsymbol m$-partite entanglement}

A state $\hat \rho$ of an $m$-partite system is called genuinely $m$-partite entangled \cite{Teh2014, Loock2003} if it cannot be represented as
\begin{equation}
 \hat \rho = \sum_\alpha P_\alpha \sum_k \eta_{\alpha,k} \,\hat \rho_{\alpha,k}\, ,
 \label{eq:biseparable}
\end{equation}
where $\alpha$ labels all possible bipartitions $(A_\alpha,B_\alpha)$ of the system. Here, $A_\alpha$ is a subset of the $m$ parties $\{1,2,\ldots m\}$ and $B_\alpha$ its complement. $\hat \rho_{\alpha,k}$ are products $\hat \rho_{\alpha,k} = \hat \rho_{A_\alpha,k}\otimes \hat \rho_{B_\alpha,k}$ of the density matrices describing the state of subsystems $A_\alpha$ and $B_\alpha$. 
The conditions $\sum_\alpha P_\alpha=1$ and $\sum_k \eta_{\alpha,k}=1$ ensure that the total density matrix is normalized. The summation over $k$ is needed since for any bipartition $\alpha$ the state can still be a statistical mixture of states that are separable with respect to this bipartition.
Our goal is to find observables, $\hat u$ and $\hat v$, and a bound $b$ such that $\Delta u \Delta v \geq b$ for all states of the form \eqref{eq:biseparable}. Thus, if this bound is violated the system must be genuinely $m$-partite entangled.

Let $\hat Q_i$, $\hat P_i$ be Hermitian operators (observables) acting on subsystem $i$. We define the observables
\begin{equation}
  \hat u=\sum_i g_i \hat Q_i \,, \qquad \hat v=\sum_i h_i \hat P_i \, ,
\end{equation}
where $g_i$ and $h_i$ are real numbers, and estimate the product of their variances
\begin{equation}
 \begin{aligned}
   \Delta^2 u \,\Delta^2 v \geq &
 \left(\sum_\alpha P_\alpha \sum_k \eta_{\alpha,k} (\Delta^2 u)_{\alpha,k} \right)
 \left(\sum_\alpha P_\alpha \sum_k \eta_{\alpha,k} (\Delta^2 v)_{\alpha,k}\right) \\
 \geq & \left(\sum_\alpha P_\alpha \sum_k \eta_{\alpha,k} (\Delta u)_{\alpha,k}(\Delta v)_{\alpha,k} \right)^2 \, .
 \label{eq:var_est1}
 \end{aligned}
\end{equation}
For the first inequality we use that the variance is concave. The second inequality is of Cauchy-Schwarz type. By $(\Delta u)_{\alpha,k}$ we denote the square root of the variance $(\Delta^2 u)_{\alpha,k}$ of $\hat u$ in state $\hat \rho_{A_\alpha,k}\otimes \hat \rho_{B_\alpha,k}$.
We now calculate a lower bound on $(\Delta u)_{\alpha,k}(\Delta v)_{\alpha,k}$ for each bipartition $\alpha$. Then the convex sum in \eqref{eq:var_est1} must be larger than the minimum one of all these bounds.

For a fixed bipartition $(A,B)$, dropping the indices $\alpha$ and $k$ and defining $\hat u_{A}=\sum_{i\in A}g_i \hat Q_i$ and  $\hat u_{B}=\sum_{i\in B}g_i \hat Q_i$ (and similarly for $\hat v$), the separability of $\hat \rho=\hat \rho_A \otimes \hat \rho_B$ implies that the covariance between observables acting on the two subsystems vanishes, and therefore 
\begin{equation}
 \Delta^2 u = \Delta^2 u_A +  \Delta^2 u_{B}
\end{equation}
and similarly for $\hat v$.
It is always true that 
\begin{equation}
 \begin{aligned}
   \Delta^2 u\, \Delta^2 v & = 
 \left[ \Delta^2 u_{A} +  \Delta^2 u_{B} \right]
 \left[ \Delta^2 v_{A} +  \Delta^2 v_{B} \right] \\
 & \geq \left[ \Delta u_{A}\Delta v_{A} + \Delta u_{B}\Delta v_{B}  \right]^2 \\
 & \geq \left[ \frac{1}{2} \left| \left\langle \left[\sum_{i\in A} g_i \hat Q_i,\sum_{i\in A} h_i \hat P_i \right] \right\rangle \right| + \frac{1}{2} \left| \left\langle \left[\sum_{i\in B} g_i \hat Q_i,\sum_{i\in B} h_i \hat P_i \right] \right\rangle \right| \right]^2 \\
 & = \left[ \frac{1}{2} \left| \sum_{i\in A} g_i h_i \langle [\hat Q_i, \hat P_i] \rangle \right| + \frac{1}{2} \left| \sum_{i\in B} g_i h_i \langle [\hat Q_i, \hat P_i] \rangle \right| \right]^2 \, .
 \end{aligned}
\end{equation}
Here we used that, for real numbers $x$ and $y$, one has $x^2+y^2 \geq 2xy$ and estimated the variance products with their lower bounds given by the Heisenberg uncertainty principle. To this point the derivation is still completely general. 

We now specify $\hat Q_i$ and $\hat P_i$ to be two non-commuting (orthogonal) quadrature operators, $\hat F(0)$ and $\hat F(\pi/2)$ (see above), such that we obtain the commutator $|\langle[\hat F_{i}(0),\hat F_{i}(\pi/2)]\rangle|=2\langle \hat N_i\rangle$ (neglecting corrections due to side-mode occupation \cite{SuppInfo}). Thus we have, for each partition $(A_\alpha,B_\alpha)$,
\begin{equation}
 (\Delta u)_\alpha (\Delta v)_\alpha \geq \left|\sum_{i\in A_\alpha}g_i h_i \langle \hat N_i\rangle\right| + \left|\sum_{i\in B_\alpha}g_i h_i \langle \hat N_i\rangle\right|\,,
 \label{eq:mult_ent_ineq}
\end{equation}
and therefore, for the sum in eq.\ \eqref{eq:var_est1},
\begin{equation}
 \Delta u \,\Delta v \geq \min_\alpha (\Delta u)_\alpha (\Delta v)_\alpha \, .
\end{equation}
Experimentally (cf.\ Fig.~\ref{Multipartite} in the main text) we measure the steering of subsystem $1$ (A) by the conjunction all other subsystems (B) and we choose $g_1=h_1=1$ as well as $g_{i>1}=g$, $h_{i>1}=h$. 
In the following we write $\langle \hat N_i\rangle=N_i$ and consider the case where the sizes of the subsystems $ N_{i>1} = (N-N_1)/(m-1)= N_{2}$ are all equal. $g$ and $h$ are determined by optimizing the steering product. With this, Eq.~\eqref{eq:mult_ent_ineq} simplifies to
\begin{equation}
  \Delta u \,\Delta v \geq \min_{k=1\ldots m-1} \biggl[ | N_1 + (k-1)g h N_2| + (m-k)|g h| N_2 \biggr]\, .
\end{equation}
We consider the case where $ N_1= \eta_{\rm A} N$ and $N_{i>1}=(1-\eta_{\rm A})N/(m-1)$. In this case $ N_1+(m-2)g h N_2>0$ is always fulfilled and thus the minimum is attained for $k=m-1$. Note that for all cases considered here $g$ and $h$ have opposite signs. We obtain the bound
\begin{equation}
  \frac{\Delta u\Delta v}{ N_1} \geq \left(1 + \frac{m-3}{m-1}g h\frac{1-\eta_{\rm A}}{\eta_{\rm A}}\right)\, .
  \label{eq:bounds}
\end{equation}
\begin{figure}
   \centering
   \includegraphics[width=0.5\textwidth]{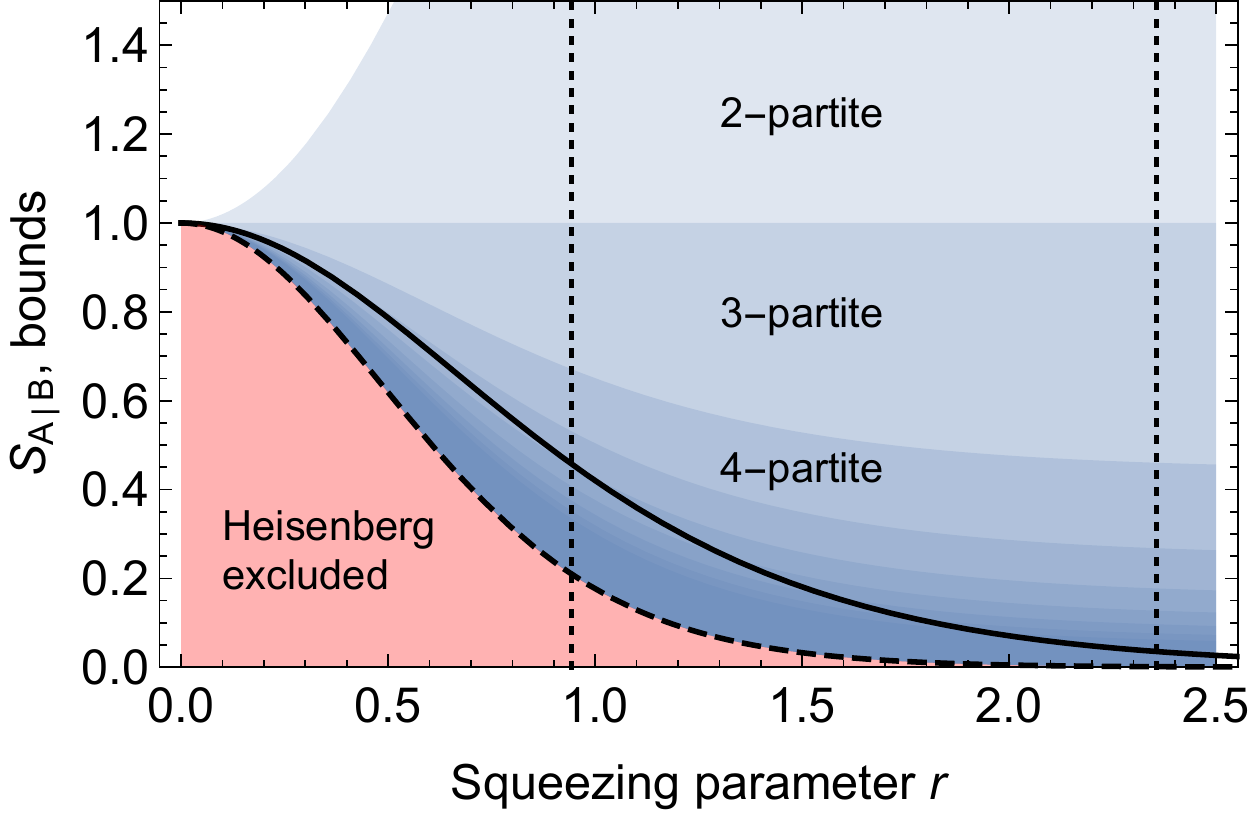}
   \caption{
   \textbf{Violation of entanglement bounds as a function of the squeezing parameter.} Splitting the cloud into A with $N_\text{A}=N/2$ and B, sub-partitioned into equal slices. The solid line is the left-hand side of eq.~\eqref{eq:boundsEtaHalf}. The blue shadings show the regions of genuine $m$-partite entanglement, cf.\ right-hand side of eq.~\eqref{eq:boundsEtaHalf}, with darker blue indicating larger $m$. The dashed line is the limit $m\rightarrow\infty$ of the bound, which coincides with the bound set by the Heisenberg uncertainty relation for the total system. The largest experimentally achieved value of $r$ is $2.3$. We note that beyond this value the undepleted-pump approximation breaks down for a total atom number of $N\approx 10^4$. The dotted vertical lines indicate the squeezing parameter corresponding to spin squeezing times $60\,$ms and $150\,$ms, respectively.}
   \label{fig:bounds1}
\end{figure}

Using the analytical results in undepleted-pump approximation derived in the previous section we evaluate both sides of the inequality and thus show how strongly the input states have to be squeezed in order to violate the bound for $m$-partite entanglement, where 
\begin{equation}
 gh=-\frac{4\eta_{\rm A}^2 \tanh^2(r)}{1-(1-2\eta_{\rm A})^2\tanh^2(r) }\,.
\end{equation}

\begin{figure}
   \centering
   \includegraphics[width=0.9\textwidth]{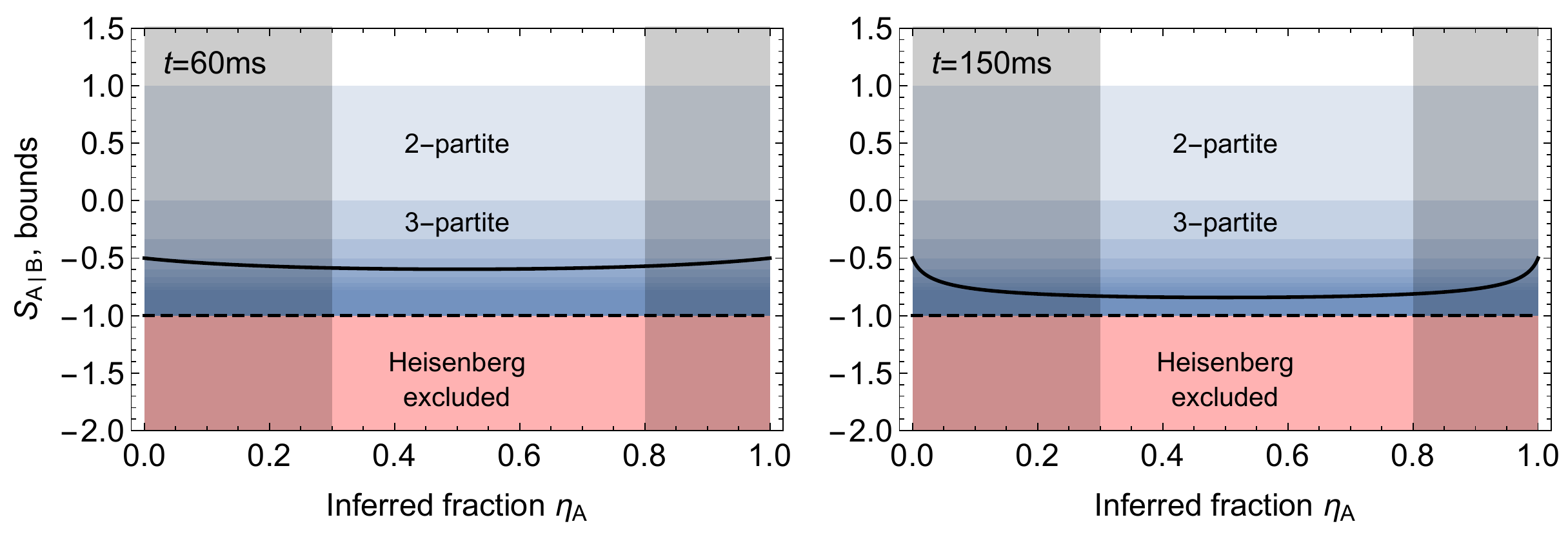}
   \caption{
   \textbf{Violation of entanglement bounds as a function of  $\boldsymbol {\eta_\text{A}}$.} Idealized case of Fig.~\ref{Multipartite} in the main text. The gray shadings show the areas that we exclude experimentally to avoid biasing due to classical correlations. Lines and shadings have the same meaning as in Fig.~\ref{fig:bounds1}.}
   \label{fig:bounds3}
\end{figure}

In the case of $\eta_{\rm A}=1/2$, i.e. $-g=h=\tanh(r)$ this gives
\begin{equation}
 \frac{\Delta u\Delta v}{N_1} \geq \left( 1-\frac{m-3}{m-1}\tanh^2 (r) \right)\, .
 \label{eq:boundsEtaHalf}
\end{equation}
The left-hand side gives $1-\tanh^2(r)$, showing that in the limit $m\rightarrow \infty$, the bound can never be violated.
Figure~\ref{fig:bounds1} illustrates these bounds for the case of $\eta_{\rm A}=1/2$. One can re-write inequality \eqref{eq:bounds} in the form given in the main text such that the bound only depends on $m$. In Fig.~\ref{fig:bounds3} we show both sides of this inequality as a function of the splitting ratio $\eta_{\rm A}$.

\newpage

\bibliographystyle{EPRKunkel_BibStyle}
\bibliography{EPRKunkel_Bib}

\end{document}